\documentclass[aps,twocolumn,prl,superscriptaddress]{revtex4-1}
\usepackage{amssymb, graphicx, hyperref}
\usepackage[intlimits]{amsmath}
\usepackage[english]{babel}
\usepackage{ulem}
\usepackage{xcolor}

\begin{document}
\title{ Bound to unbound states transitions of heavy quarkonia in the cooling phase of QGP}

\author{Nirupam Dutta}
\email[E-mail:]{nirupamdu@gmail.com}
\affiliation{School of Physical Sciences, National Institute of Science Education and Research Bhubaneswar, HBNI,
P.O. Jatni, Khurda 752050, Odisha, India}

 \author{Partha Bagchi}
\email[E-mail:]{p.bagchi@vecc.gov.in, parphy85@gmail.com}
\affiliation{Theoretical Physics Division, Variable Energy Cyclotron Centre,
1/AF, Bidhan Nagar, 
Kolkata 700064, India}\date{\today}
\author{Jobin Sebastian}
\email[E-mail:]{jobinsk93@gmail.com, jobin.sebastian@niser.ac.in}
\affiliation{School of Physical Sciences, National Institute of Science Education and Research Bhubaneswar, HBNI,
P.O. Jatni, Khurda 752050, Odisha, India}

\begin{abstract}
Emphasizing the possibility of moderate suppression of heavy quarkonium states, we invite some attention towards the issue of real time evolution of quarkonia during the cooling phase of quark gluon plasma(QGP). In this context, we have used time dependent perturbation theory to show  that $\Upsilon(1S)$,  $\Upsilon(2S)$, $J/\Psi$ can further be dissociated in the medium at a temperature below their dissociation thresholds even though they survive the Debye screening. We have presented and compared the dissociation probabilities and dissociation rates of these states in real and complex valued potential in this article. We realise that our method is an approximate way to analyse the short time behaviour of quarkonium in real valued potential and for long time behaviour one must adopt a non perturbative technique for solving schr\"odinger equation of quarkonium bound states in evolving QGP. On the other hand the perturbation technique seems to be valid enough to deal with the same for a complex valued quark anti-quark potential.
\end{abstract}

\maketitle
In recent years various studies \cite{akamatsu2015, rothkopf14,dusling13,brambilla17,katz15,deboni17} on heavy quarkonia have been directed towards understanding their real time dynamics in static and evolving quark gluon plasma (QGP). These approaches either rely on the modelling of heavy quarkonia as open quantum systems or on the solution of schr\"odinger equation in the potential model. In the evolving quark gluon plasma (QGP), the quantum dynamics of quarkonia is important as the energy eigenstates do not remain the same eigenstates of the time dependent Hamiltonian. This is the essence of non-adiabatic evolution \cite{Dutta:2012nw}, which becomes relevant in a rapidly evolving medium. Furthermore, for a static QGP, calculations show \cite{borghini2011,akamatsu2011} that the energy eigenstates can make transitions to other states due to the medium interaction. The medium interaction can cause suppression of quarkonium species through transition from one bound state to other bound states (reshuffling) and unbound states (dissociation). The survival or dissociation mechanism is also important for understanding the anisotropic flow of different quarkonium states \cite{das18,bhaduri18}. In earlier works \cite{borghini2011,akamatsu2011}, the bound states to bound states transitions have been addressed in a static medium in the open quantum system model by solving master equations or stochastic schr\"odinger equation but the dissociation probability has not been calculated so far. In recent times the survival probabilities of various states have been calculated \cite {brambilla17} by using tools and techniques of open quantum systems. In dynamical picture, if we start with a specific ground state population, some of them are going to be dissociated completely which contributes to the enhancement of quarks ($q$) and antiquarks ($\bar{q}$) in the medium and another part of the population is subjected to transitions to other possible bound states.  In potential model the transitions can happen due to the change in the quark anti quark potential caused by the change of medium temperature. As far as bound states to bound states transitions are concerned, it can happen differently in different temperature regimes of the medium evolution. Let us demonstrate this issue in a bit more detail. Suppose we have the ground states of charmonia and bottomonia at the beginning of  QGP formation. Now as the medium evolves, quarkonia will make transitions  to excited states which cannot be stable unless the medium temperature drops below the dissociation temperatures of those excited states. Bottomonium ground states can survive even above $3T_c$ ($T_c $ being the pseudo critical temperature of QCD phase transition taken to be $155$MeV for this article) and $\Upsilon(2S)$ can not survive above $1.5T_c$ approximately. There have been several works done in this context for determining the dissociation temperature over the years, which have quoted different values \cite{ding12,karsch05,aarts10,aarts11,aarts12,mocsy07}. For this article we have considered the above mentioned dissociation temperature of the first excited state of bottomonia. So, during the phase when medium temperature falls off from $3T_c$ to $1.5T_c$ (in LHC scenario) there will be only dissociation from the ground state $\Upsilon$(1s) but no transition to excited states. Obviously after that when QGP temperature decreases further up to $T_c$, the ground state can make transition to $\Upsilon(2S)$ as well as to the unbound states. In this phase the first excited state $\Upsilon(2S)$ can make transition to the ground state as well as to the unbound states. So the dissociation of $\Upsilon(1S)$ happens throughout the medium evolution but the transition to first excited state becomes important after the medium temperature reaches below $1.5 T_c$.

Similarly, for charmonium states, we cannot expect any bound states at $3T_c$ but can expect $J/\Psi$ at below $1.7T_c$ if they are formed late in the medium through feed down or regeneration of quarks and antiquarks \cite{du15}. Thus, when the medium temperature falls from $1.7T_c$ to $T_c$, we can expect transitions to unbound states only as no other states are permitted to survive above $T_c$ for charmonia  before the statistical hadronization \cite{andronic05}. So, for charmonia we are interested in calculating the dissociation of $J/\Psi$ during this phase of cooling. It is obvious from the discussion we have made that the dissociation from different bound states happen sequentially during the course of medium evolution.  We should remember that here we are alluding to the dissociation through the real time evolution of quarkonia which is completely different from that due to the screening of the plasma or gluodissociation process. To the best of our knowledge this has not been addressed till date. An attempt \cite{bagchi2014} was made earlier in this direction to understand the dissociation during thermalization by using sudden approximation. In this article, we will address this issue of sequential dissociation over the time span of medium evolution. It is worth mentioning that the calculation of transition probabilities to bound and unbound states is cumbersome as one has to solve the Schr\"{o}dinger equation for a time dependent potential. 

We adopt time dependent perturbation theory to calculate the dissociation probability in potential model. 
The heavy quark potential in medium can be written as,
\begin{equation}
V(r) =  - \frac{\alpha}{r} exp(-m_D r) +
        \frac{\sigma}{ m_D} (1 - exp(-m_D r)) \label{realpotential}
\end{equation}

where, strong coupling ${\alpha}$ and the string tension $\sigma$ are taken to be around $0.47$ and $0.2$ respectively. Obviously these constants change with temperature. In the evolving plasma they are therefore varying with time but small compared to the variation of the quantity Debye mass $m_D$ over the span of medium evolution. Essentially, in our calculation the Debye mass is making the potential as a time dependent quantity in the QGP whose temperature is falling off in the following way \cite{blaizot90},

 \begin{equation}
T(t) =T_0\big(\frac{\tau_0}{\tau_0+t} \big)^{\frac{1}{3}}\label{bjorken}.
\end{equation}
$T_0$, the initial temperature and $\tau_0$ be the thermalization time of QGP and taken to be approximately $0.5$fm/c in our calculation. At the beginning ($t=0$), just after the thermalization of the medium, at a temperature $\approx 500$MeV, the only possible state which can persist in the medium is the ground state of Bottomonia. The other states are already dissociated due to color screening of the medium at that temperature. We are interested in calculating the dissociation probability of $\Upsilon(1S)$ during the entire cooling phase of QGP whereas for charmonia we will be calculating the dissociation probability of $J/\Psi$ which evolves during the phase when medium temperature falls off from $1.7T_c$ to $T_c$. In case of  $\Upsilon(2S)$, we will be calculating the same for the medium temperature $1.5T_c$ to $T_c$. We should definitely mention here that though these excited states are not expected to be present in the beginning, they can appear through the recombination of quark antiquark pairs available in the medium or through the non adiabatic transitions from the ground state \cite{Dutta:2012nw,Das:2018xel}.

The time dependent perturbation at any instant $t$ can be considered as $H^1(t)=V(r,t)-V(r,0)$. It is obvious that the instantaneous eigenstates of the time dependent Hamiltonian are changing with time but we can consider upto a reasonable level of approximation that the unbound states are the plane wave states lying in the continuum regime of energy. Hence they can be expressed as,
\begin{equation}
\Psi_{k} =\frac{1}{\sqrt{\Omega}}e^{i\vec{k}.\vec{r}}
\end{equation}
normalised over a volume $\Omega$ and can have different possible values of the momentum $\vec{k}$ in the continuum regime of momentum. Suppose the heavy quarkonia is initially in the eigenstate $ |\Psi_i\rangle$ of the potential $V(r,0)$. Now, after time $t$ the first order contribution to the transition amplitude to the plane wave state characterised by the momentum $k$ is given by,
\begin{equation}
 a_{ik}=\int\frac{d}{dt}\langle \Psi_k|H^1(t)|\Psi_i\rangle\frac{e^{i(E_i-E_k)t}}{(E_i-E_k)} dt \label{amp}.
\end{equation}
We should remember that in this case, the perturbation is never switched off rather it always maintain non zero value for the whole span of time we have considered. For a rigorous understanding of such processes one can look into the relevant discussion made in the book by L. Landau \cite{landau1958course}.

Here, $E_i$ is energy eigenstate corresponding to the initial state $|\Psi_i\rangle$ and  $|\Psi_k\rangle$ is the unbound state with the momentum $k$. The bound state can make transition to all possible momentum states and the total probability of dissociation is therefore the sum of the transition probabilities to all possible unbound states designated by the continuum span of momenta ranging from zero to infinity.

The number of unbound states between the momentum continuum $k$ and $k+dk$ over  $4\pi$ solid angle is given by,
\begin{equation}
 dn=\left(\frac{L}{2\pi}\right)^3 k^2 dk =\frac{\Omega}{(2\pi)^3} k^2 dk \label{density},
\end{equation}
where $L$ is the size of the box for the normalization of plane wave states.The total transition probability to all continuum states is given by,
\begin{equation}
 P=\int_{k=0}^{\infty}|a_{ik}|^2\frac{\Omega}{(2\pi)^3} k^2 dk .
\end{equation}

Binding energies and wave functions of bound states have been calculated by using Numerov method. For an optimum performance of our numerical calculation, we have considered the initial temperature of the thermalised deconfined medium at $t=0$ to be around $400 $MeV. We have considered the charm and bottom mass as $1.3$GeV and $4.66$GeV respectively for our calculation.

 So far, we have dealt with the bound state to unbound state transitions of different quarkonium states evolving in a time dependent real valued potential. In recent times it has been realised that the quarkonium states in medium should be described in potential model through a complex valued potential \cite{laine06,brambilla17,mike11,kajimoto17}. Effective theories or modelling in open quantum system has indicated that the effective potential of a small system interacting with many degrees freedom can actually acquire an imaginary part. This manifestation of Landau damping makes a fundamental change in the description of quarkonium evolution as the states are no longer being characterised by stationary states. Rather the non-unitary evolution introduces a finite life time for them. This eventually  causes a dynamical melting through the broadening of the width of the bound state with time. This is obviously true even in the static QGP which previously has been investigated. In this article we are focusing on non-adiabatic transitions which becomes unavoidable in an evolving deconfined medium. Now the relevant question is that if such an imaginary potential is considered, does it make any drastic difference to the bound states to unbound states transitions. To investigate this, we consider an imaginary part in the potential evaluated in \cite{laine06,mike11} and the initial unmelted quarkonium states ($\Upsilon(1S)$, $J/\Psi$, $\Upsilon(2S)$ created and existed in various phases of QGP stated earlier in this article), as bound states (even though they may have some decay width) and then by applying the rule of first order perturbation theory , we have the the following estimates of transition probabilities and rates.
 
 We have considered an additional imaginary part with the real valued potential quoted earlier in this article in eq.\ref{realpotential}. Hence the complex potential is given by,
 
 \begin{equation}
V_{com}(r) = V(r)+i (-\alpha T\Phi(m_{D}r)),\label{complex}
\end{equation}
where $\Phi$ a dimensionless function defined as is defined in the following way,

\begin{equation}
\Phi(x)=2\int_{0}^{\infty}dz\frac{z}{(z^2+1)^2}\big(1-\frac{\sin(zx)}{zx}\big)\label{regularization}.
\end{equation}

Now with the inclusion of the imaginary part in the potential the expression for the transition amplitude becomes,
\begin{equation}
 \bar{a}_{ik}=\int_0^{t}\frac{d}{dt}\langle \Psi_k|H_{com}^1(t)|\Psi_i\rangle\frac{e^{i(E_i-E_k)t}}{(E_i-E_k)-i\frac{\Gamma}{2}} e^{\frac{\Gamma}{2} t}dt \label{amp},
\end{equation}
\begin{figure}
\includegraphics[width=0.5\textwidth]{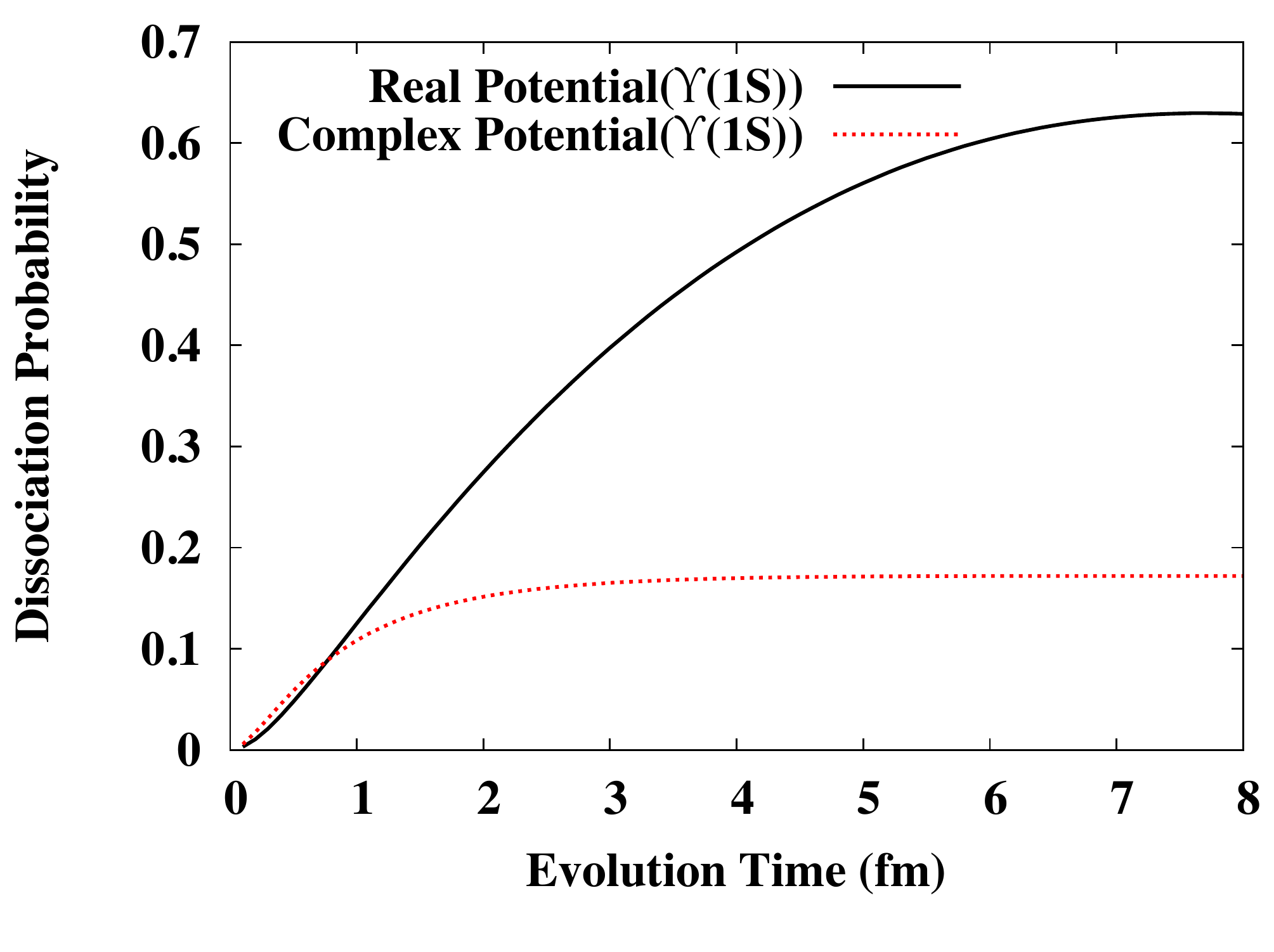}
\caption{\small The solid black line shows that dissociation probability (transition to unbound states) of $\Upsilon(1S)$ increases with time (obeying the Fermi golden rule) in real valued potential and the dotted red curve shows the same in the complex valued potential.}
\label{fig:one}
\end{figure}

\begin{figure}
\includegraphics[width=0.5\textwidth]{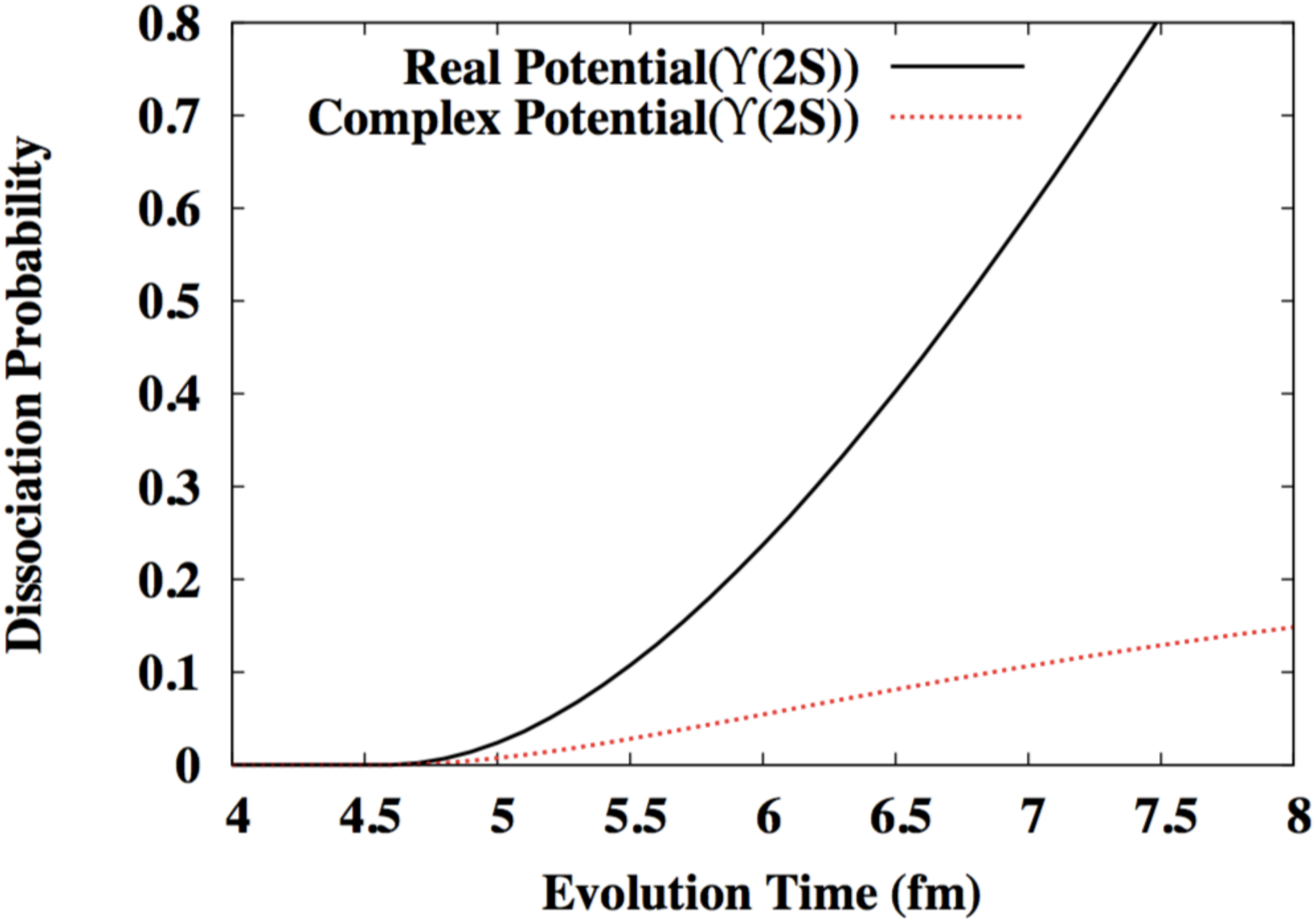}
\caption{\small The solid (black) and dotted (red) lines show the dissociation probabilities of $\Upsilon(2S)$ below 1.5 $T_c$ as a function of time in real and complex valued heavy quark potential respectively.}
\label{fig:two}
\end{figure}
\begin{figure}
\includegraphics[width=0.5\textwidth]{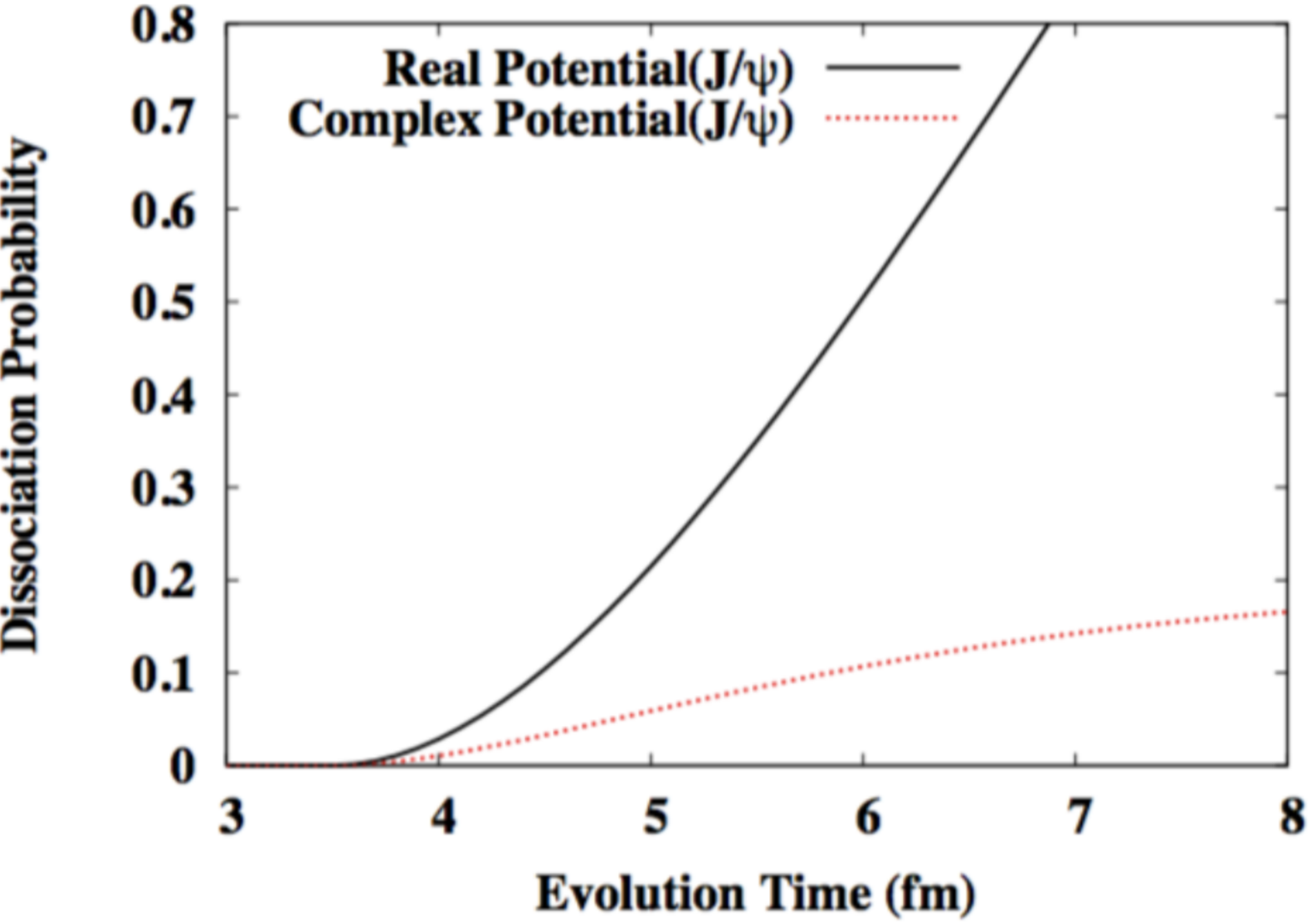}
\caption{\small he solid (black) and dotted (red) lines show the dissociation probabilities of $J/\Psi$ below 1.7 $T_c$ as a function of time in real and complex valued heavy quark potential respectively.}
\label{fig:three}
\end{figure}
with $H_{com}^1(t)$ as the perturbation in complex potential. The additional quantity $\Gamma$ appears due to the imaginary part in the potential $V_{im}(r)$ and is proportional to the imaginary contribution to the energy of the initial bound state \cite{kajimoto17}  which can be estimated through the following equation,
  \begin{equation}
  \Gamma \approx 2 \int dr V_{im}(r) |\Psi_{i}(r)|^2 . \label{gamma}
  \end{equation}
  It can be checked that the factor $\Gamma$ is a a negative quantity which shows the decay of the bound state. While calculating the transition amplitude, the initial state can always be normalised by the wave function corresponding to the real part of the potential without any loss of generality \cite{kajimoto17,deboni17}and we can introduce the medium interaction at later time. This is obvious as the decoherence time is the scale above which the non unitary open system dynamics sets in.
  \begin{figure}
\includegraphics[width=0.5\textwidth]{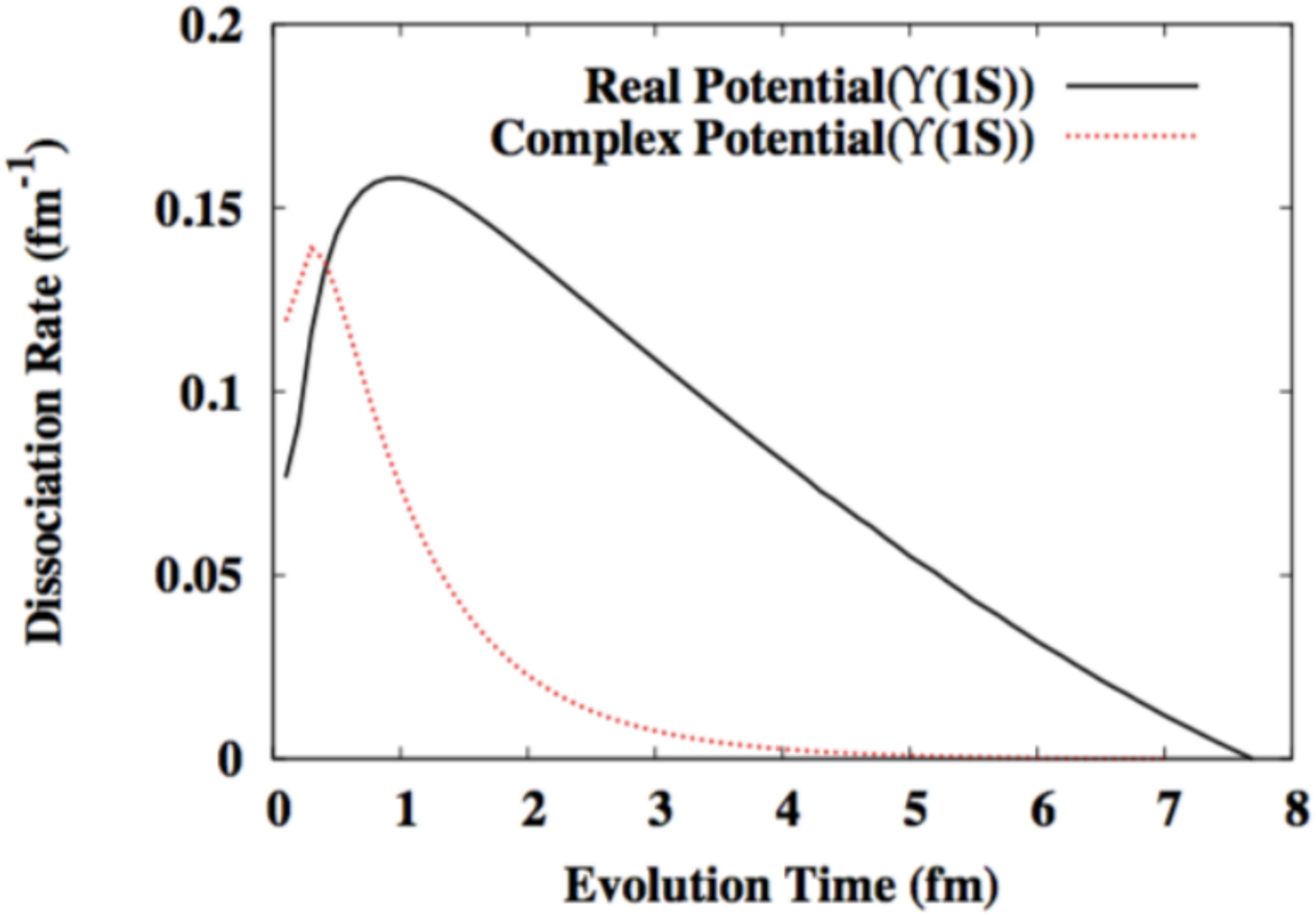}
\caption{\small The solid (black) and dotted (red) lines show that transition rate of $\Upsilon(1S)$ to unbound states with time in real and complex valued potential respectively.}
\label{figure:four}
\end{figure}

\begin{figure}
\includegraphics[width=0.5\textwidth]{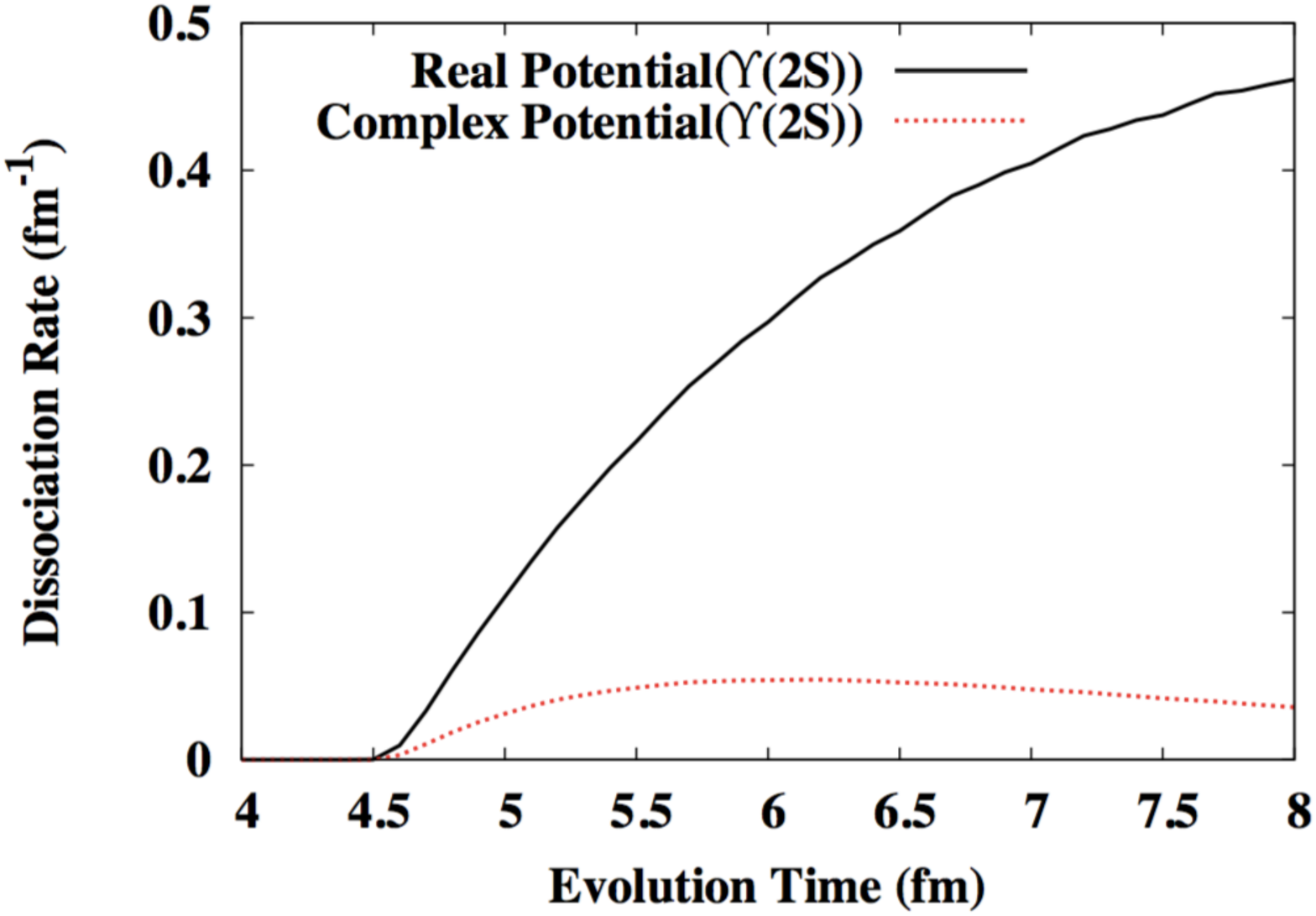}
\caption{\small The solid (black) and dotted (red) lines show that transition rate of $\Upsilon(2S)$ to unbound states with time in real and complex valued potential respectively.}
\label{figure:five}
\end{figure}

\begin{figure}
\includegraphics[width=0.5\textwidth]{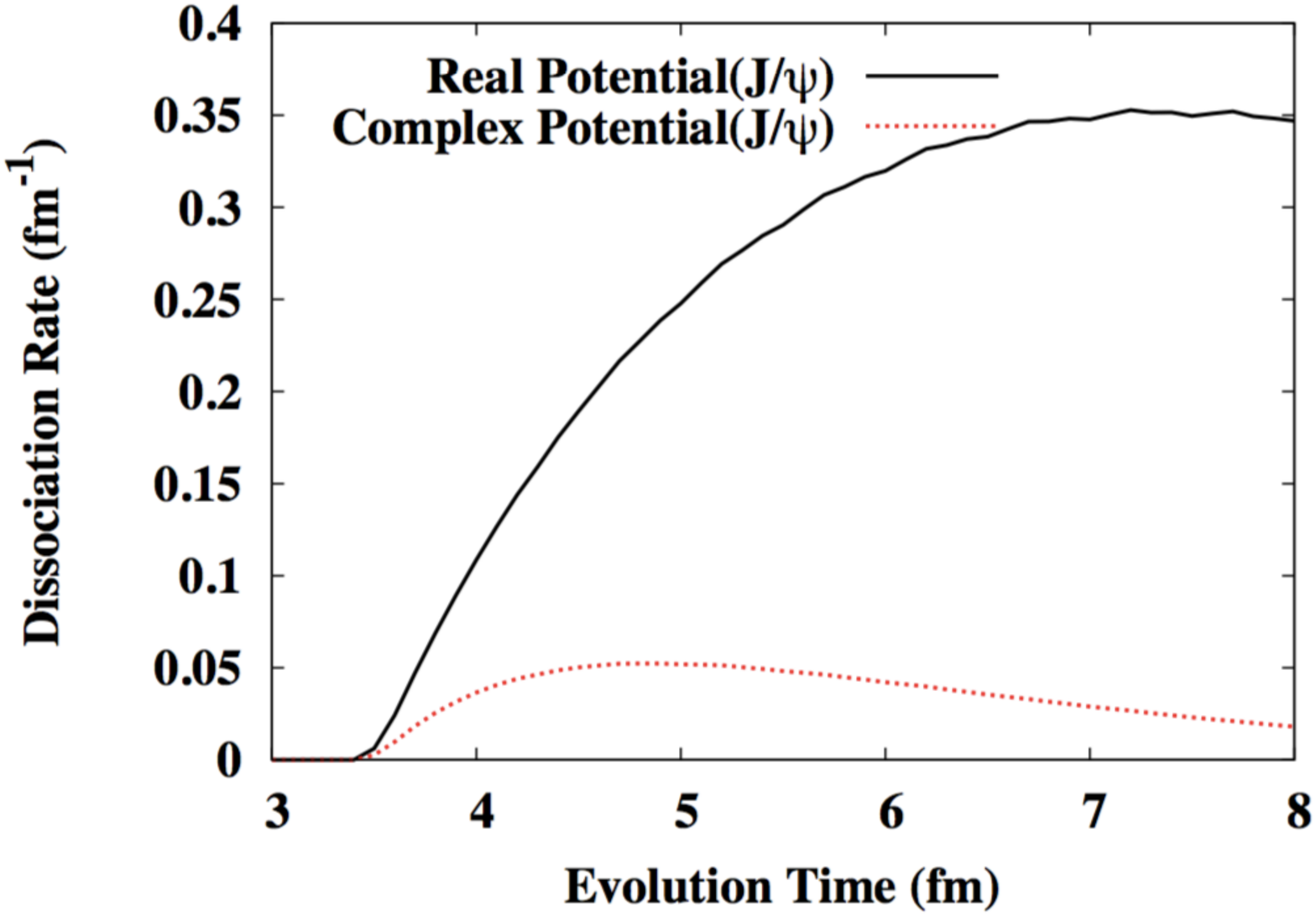}
\caption{\small The solid (black) and dotted (red) lines show that transition rate of $J/\Psi$ to unbound states with time in real and complex valued potential respectively.}
\label{figure:six}
\end{figure}

We already have discussed that the dissociation of $\Upsilon(1S)$ can occur during the complete cooling phase of QGP. In figure \ref{fig:one}, we have presented the comparison between the dissociation probability of $\Upsilon(1S)$ (transition to unbound states) in real (solid black curve) and complex potential (dotted red curve) respectively. We see that for real valued potential the dissociation probability of  $\Upsilon(1S)$ increases almost linearly with time. This is quite understandable from the point of view of Fermi golden rule in first order perturbation theory. It is obvious that the first order perturbation theory is not sufficient to account for the dissociation probability when it is not moderately less than $1$. For that reason we can apply the technique for not more than few fm in case of real valued potential. Nevertheless, it shows the very physical fact that, due to the rapid change of potential there must be dissociation of  relevant quarkonium state which one needs to consider while addressing the suppression of $\Upsilon(1S)$ in an evolving QGP. In the complex valued potential, the transition probability to unbound states does not increase very rapidly as the decay of bound state plays an important role. The dissociation rate shows faster decay in comparison to the one in real valued potential as the survival of bound states are now convoluted with additional exponential decay factor due the appearance of imaginary part in the potential. This decay in transition rate for $\Upsilon(1S)$ is expected to be greater than the decay for other two states as $\Upsilon(1S)$ spends longer time in the medium and this is clear from the plot given in fig. \ref{figure:four}. The dissociation rates are much higher in the beginning and later it decreases. This is quite expected as the temperature of the medium initially falls off very rapidly and later the change in the temperature becomes small causing comparatively less rate of change of potential. We know that higher the change of potential, higher is the probability of transitions to other states.  The medium cools down to a temperature $230$MeV after a time around $2$fm. Below this temperature, the $\Upsilon(2S)$ state can survive the Debye screening and they can be produced in the medium even though the the medium does not allow them to be created initially in the very hot atmosphere. This late production of $\Upsilon(2S)$ may happen only through the non-adiabatic transitions from the ground state itself. They are less likely to be created through recombination of $b$ and $\bar{b}$ pairs in the medium as the number of such pairs is very small. If the excited state is formed, there is certain probability of their dissociation again due to the medium evolution. Similar argument applies to the case of $J/\Psi$ which can be formed due to the recombination of $c$ and $\bar{c}$ pairs only when the medium temperature drops down to $1.7T_c$ after a time around $1.24$fm. In fig. \ref{fig:two} and \ref{fig:three} we have plotted the transition probabilities to continuum in real and complex valued potential for $\Upsilon(2S)$ and $J/\Psi$ respectively. We observe that the dissociation rates in real potential of these two states are higher than that of $\Upsilon(1S)$ and it is simply because they are so loosely bound that they can be dissociated heavily even though the rate of change of potential in the later stage is small. In this case also we see the glimpse of Fermi golden rule and the results can be taken seriously up to around $5-6$fm only for real potential. In the complex potential scenario the application of perturbation theory becomes quite valid almost for the complete span of cooling phase. The transition rates to unbound states for  $\Upsilon(2S)$ and $J/\Psi$ are presented in fig. \ref{figure:five} and fig. \ref{figure:six} respectively.



In conclusion, we say that the initially survived $\Upsilon(1S)$ and later produced $\Upsilon(2S)$ and $J/\Psi$ can further be dissociated in the medium at a temperature even below their respective dissociation thresholds. The medium induced transition was previously considered to study the bound state to bound state transition for a static QGP \cite{borghini2011} using the solution of master equation but the transition to unbound states was untreated. In this article we have addressed the problem of bound state to unbound state transitions and that also for a more realistic scenario of evolving medium. Our calculation in real valued potential gives an approximate result which is valid to describe short time behaviour. For long time behaviour, one must look into a non perturbative computation which at the moment does not appear feasible. On the other hand, in the complex potential scenario, the additional decay governed by the imaginary part makes the perturbation technique extremely reliable almost for the whole time span of the cooling phase of QGP.

\begin{acknowledgements}
N. D. acknowledge support from theoretical high energy physics department, University of Bielefeld during the initial stage of this research. P. B. acknowledges SERB (NPDF Scheme: PDF/2016/003837), Government of India for the financial assistance. J. S. acknowledges support  from School Of Physical Sciences (SPS), NISER.
\end{acknowledgements}

\bibliography{th}{}
\end{document}